\documentclass[showpacs,twocolumn,preprintnumbers,amsmath,amssymb]{revtex4-1}
\newcommand{\beq}{\begin{equation}}
\newcommand{\eeq}{\end{equation}}

\newcommand{\id}
 {i\kern.06em\hbox{\raise.25ex\hbox{$/$}\kern-.60em$\partial$}}

\newcommand{\bs}{/\kern-.52em b}
\newcommand{\qs}{/\kern-.52em s}

\newcommand{\p}{\partial}

\newcommand{\dd}
{\kern.06em\hbox{\raise.25ex\hbox{$/$}\kern-.60em$\partial$}}

\newcommand{\lj}{\langle}
\newcommand{\rj}{\rangle}
\newcommand{\vep}{\varepsilon}
\newcommand{\bx}{\boldsymbol{x}}
\newcommand{\by}{\boldsymbol{y}}
\newcommand{\bk}{\boldsymbol{k}}

\newcommand{\bz}{\boldsymbol{z}}
\newcommand{\bp}{\boldsymbol{p}}

\newcommand{\bP}{\boldsymbol{P}}

\newcommand{\uk}{\underline{k}}

\newcommand{\bO}{{\bf O}}

\newcommand{\mKG}{m_{ \text{KG}}}

\usepackage{graphicx}
\usepackage{dcolumn}
\usepackage{bm}
\usepackage{mathrsfs}

\usepackage{multirow}
\usepackage{empheq}
\usepackage[theorems,skins]{tcolorbox}

\DeclareMathAlphabet{\mathpzc}{OT1}{pzc}{m}{it}

\input amssym.def
\input amssym

\input epsf

\begin{document}
\title{A  Generally Covariant Theory of  Quantized Real Klein-Gordon Field  in de Sitter Spacetime}
\author{Sze-Shiang Feng}
\address{9521 Windwood Point, Dayton, Ohio 45458, USA,\\
 email:sshfeng2014@gmail.com}
\date{\today}

\baselineskip 0.17in
\begin{abstract}
We propose in this paper a quantization scheme for real Klein-Gordon field in de Sitter spacetime.  Our scheme is generally covariant with the help of vierbein, which is necessary usually for spinor field in curved spacetime. We first present a Hamiltonian structure, then quantize the field following the standard approach.  For the free field, the time-dependent quantized Hamiltonian is diagonalized by Bogliubov transformation and the eigen-states at each instant are interpreted as the observed particle states at that instant. The interpretation is supported by the known cosmological red-shift formula and the on-shell condition of 4-momentum for a free field. Though the mathematics is carried out in term of conformal coordinates for the sake of convenience, the whole theory can be transformed into any other coordinates based on general covariance.  It is concluded that particle states, such as vacuum states in particular are time-dependent and vacuum states at one time evolves into non-vacuum states at later times.  Formalism of perturbational  is provided with an extended Dirac picture.
\end{abstract}

\maketitle

\section{Introduction}
Is it necessary to pursue a quantum field theory conforming to the principle of general covariance and why another discussion on the same topic? To the first question, 
 the answer seems affirmative. To answer the second, one needs to evaluate lots of schemes that have been proposed as of today. It seems to the present author that most of discussions are not as physically fruitful as conventional quantum field theories, though various mathematical structures have been revealed.  And the topic of quantized scalar field in curved spacetime deserves revisits. 
 
 A unified theory that can explain all forces in nature in quantized fashion has been a Holly Grail in physics for quite long a time ever since Einstein. Yet, definite progress in relating to the real world is still called for considering the developments as of today in superstring/brane theories. In stead of seeking a final theory in which all fields in nature is quantized, a less ambitious endeavor has been paid to the quantization of all fields except gravity in curved spacetime, a branch that has been one of the major foci in theoretical physics for decades\cite{Birrell}-\cite{Bar}.   To understand the necessity of quantization of matter fields in curved spacetime, one can consider a basic question: pions in cosmic rays come down to the earth all the way from distant universe, are they quantized particles when then pass some region which maybe strongly curved by gravity? The answer is seemingly affirmative, i.e., we should have a complete theory of quantum field theory in curved spacetime.
Due to the curvature of spacetime, the canonical quantization of fields is not generally as applicable in curved spacetime as in Minkowski spacetime.  Because of this, quantization is implemented in many cases by mode expansion directly and bypass the discussion of canonical structures.  This inevitably entails the difficulty of interpretation of concepts such as {\it particles} and {\it vacuum states}.
Henceforth, observable quantities such as energy and momentum et al  are not clearly defined as in conventional Minkowski spacetime quantum field theories.  Since we can choose any coordinates system and obtain a different set of mode solutions, we need to verify the general covariance as required by general relativity itself. As in the quantization of Yang-Mills fields, one can work in different gauge conditions. But the whole framework should prove to be gauge-independent at the end of the day. Yet, general covariance is  either not proved or neglected in the various quantization schemes of matter fields in curved spacetime. 
    
     Another important issue in quantization of matter fields in curved spacetime is the specification of Fock space which represents states of quantized particles.
In conventional quantum field theories in Minkowski spacetime, one of the axioms of the LSZ framework of quantum field theory (QFT) is $P_a|0\rj=0$\cite{Bjorken} where $P_a$ is the total energy-momentum operator. Without this condition, we can draw absurd conclusions. Suppose we have Heisenberg algebra $[a,a^\dag]=1$, we can have infinite number of ways of implementing Bogliubov transformations like $a=u\alpha+v\beta^\dag, a^\dag=u^*\alpha^\dag+v^*\beta$ with $|u|^2-|v|^1=1$. But for a quadratic Hamiltonian, only one transformation can diagonalize the Hamiltonian and the observed energy quanta is represented by the creation/annihilation operators. Just as in BCS theory of superconductivity, the basic observed quanta are the quasi-particles. Nevertheless, various approaches to quantization scheme of matter fields in curved spacetime lack or neglict the Hamiltonian.

    We propose in this paper a quantization scheme for real Klein-Gordon field in de Sitter spacetime. Quantum field theories in de Sitter spacetimes have been discussed in various ways (for a survey see \cite{Gazeau2006}). To the present author, the merit lies in a number of aspects. First, Minkowski spacetime is not a solution to the Einstein equation in the presence of non-vanishing cosmological constant and the simplest solution is de Sitter spacetime. Second, de Sitter spacetime is a maximally symmetric spacetime as is Minkowski spacetime. Though it is widely recognized that discussions regarding Dirac field in curved spacetime require vierbein representing gravitational field, we here also introduce vierbein even for the quantization of Klein-Gordon field which is a scalar. The rationale behind is of three folds.  First, it is necessary even in quantizing a 1D mechanical system. Consider a system of $H=(1/2m)a^2(s)\dot{q}^2(s)+V(q)$. By re-defining time variable as $t=\int a^{-1}(s)ds$, the rest of the procedure becomes standard. Though this example seems trivial, it indicates that vierbein seems indespensible in quantizing systems in curved spacetime.  Second, we intend to put the quantization of both scalar fields and spinor fields on the same footing and keep the whole framework coordinate independent. Third, we believe that quantization is about physical observables, as Heisenberg had realized some nine decades before ( a historical account for Heisenberg's original thought is provided in\cite{Wu}).  As is well-known, Noether's  theorem reveals the intrinsic relations between conservation laws  and symmetry/invariances of  of the physical system under consideration.  Energy-momentum conservation corresponds to invariance under spacetime translation whereas angular-momentum conservation corresponds to invariance under spacetime rotation.  Yet, for general curved coordinate $x^\mu$, the invariance under transformation with $\delta x^\mu=\text{Const}$ does not correspond to {\it translation} in general since it might realize a rotation should it be an angular coordinate. Nevertheless, $\delta x^\mu=e^\mu_a b^a$  with $b^a=\text{Const}$ always represents a local spacetime translation since the projection $\delta x^\mu e^a_\mu$ of $\delta x^\mu$ on local frame $e^a_\mu$ is a translation.  This argument led to a generally covariant formulation of energy-momentum conservation of matter-gravitation system\cite{Duan1963} , a quintessential example showing the significance of vierbein in general relativity. As has been discussed in \cite{Feng1997}-\cite{Hobson}, {\it observed time and space intervals are projections of coordinate intervals onto local Lorentz frame of the observer}. The varying rate of a field in space and time should be measured over  the observed space and time instead of the coordinates. Hence, vierbein is essentially necessary  to discussions of all kind of fields. 
  
    Discussions of quantum mechanics in de Sitter spacetime was initiated shortly after the birth of relativistic mechanics of electrons\cite{Dirac1935}- \cite{Fang1980} and has been long since an important topic of quantum theories in curved spacetime.  Quantum field theories have been formulated in different approaches and different coordinate systems.  Group-theoretic approaches to quantum field theories in de Sitter spacetime are proposed in \cite{Angelopoulos1981}- \cite{Joung2007} .  Since de Sitter spacetime can be imbedded in  1+4 dimensional pseudo-Euclidean spacetime, quantum field theories have been formulated  in terms of 1+4 pseudo-Euclidean coordinates\cite{Bros1994}-\cite{Pol'shin2000}; in terms of spherical coordinates\cite{Chernikov1968}-\cite{Tagirov1973}; in terms of static coordinate \cite{Otchik1985}-\cite{Redkov2011} and in terms of co-moving coordinates   \cite{Nachtmann1967}-\cite{Casher2011}.  Unlike standard quantum field theory in Minkowski spacetime which are Lorentz invariant, most of these theories are short of either general covariance or important concepts such as Hamiltonian and measurable particle states.  In our present paper, we seek a generally invariant formalism for quantization and develop physical concepts such as particles and vacuum.   
    
   Our goals of this paper is of three folds: (i) providing a generally covariant quantum theory of Klein-Gordon field in de Sitter spacetime, in light of the fact that existing theories are short of general covariance either implicitly or explicitly; (ii) providing observable quantities of the field quanta; (ii) providing calculation approaches for scattering matrix; 
    
    The present paper is arranged as follows.  In section II, we present the canonical structure of a real Klein-Gordon field in de Sitter spacetime following the standard approach. Upon redefining canonical variables, the Hamiltonian equations of motion of canonical momentum to the field is simplified. In section III,, the system is quantized in Schr\"{o}dinger picture. Section IV is a review of fundamental solutions of Klein-Gordon field in de Sitter spacetime, as a preparation of second quantization. Section V presents in detail the field 2nd quantization and the quantized Hamiltonian both in Heisenberg picture and Schr\"{o}dinger picture. The Hamiltonians are diagonalized and quasi-particle creation/annihilation operators are defined. Discussions of difference as well as thing in common in the two pictures are provided. Time dependent vacuum and particles states are defined. Particularly, the observed energy-momentum is obtained based on our previous work and the on-shell relation for free particles is obtained. Some simple matrix elements are calculated.  In section VI, we define a generation functional which can be used to calculate various matrix elements. In section VII, transition amplitude between two states at different times is formulated. Section VIII is devoted to formulation of perturbation theory for interacting field, with the help of Dirac picture.  The last section IX is conclusional discussion and prospect of this work.

\section{Canonical Quantization of Real Klein-Gordon Field }
\subsection{de Sitter Spacetime}
\indent As a special  case of  Robertson-Walker spacetime, the de Sitter  spacetime is most easily represented as the hyperboloid\cite{Birrell}
\beq
\eta_{AB}z^Az^B=(z^0)^2-(z^1)^2-(z^2)^2-(z^3)^2-(z^4)^2=-\ell^2  \label{dSeq}
\eeq
embedded in 5-dimensional Minkowski space with metric
\beq
ds_5^2=(dz^0)^2-(dz^1)^2-(dz^2)^2-(dz^3)^2-(dz^4)^2
\eeq
(The relation between the parameter $\ell$ and the cosmological constant $\Lambda$
is $\ell=\sqrt{3/\Lambda}$. )
Choosing the coordinates $(t,\bx)$  defined by
\beq
\left\{\begin{array}{cl}
   z^0=\ell{\rm sinh}\frac{t}{\ell}+\frac{1}{2\ell}e^{t/\ell}|\boldsymbol{x}|^2\\
z^4=\ell{\rm cosh}\frac{t}{\ell}-\frac{1}{2\ell}e^{t/\ell}|\boldsymbol{x}|^2\\
z^i=e^{t/\ell}x^i
         \end{array}\right.  \label{Trans1}
\eeq
We consider the range covered by $0\leq t<+\infty, -\infty<x^i<\infty$ since we take $t=0$ as the inception of the evolution of the universe.
The induced line element on the hyperboloid is
\beq
ds^2=dt^2-e^{2t/\ell}\sum_{i=1}^3(dx^i)^2.
\eeq
Now define a conformal time
\beq
\zeta=\ell e^{-t/\ell}, \,\,\,\,\,\,\,\,\,\, 0< \zeta<\ell
\eeq
then
\beq
ds^2=C(\zeta)[d\zeta^2-\sum_i (dx^i)^2]
\eeq
with conformal factor $C(\zeta)=(\ell/\zeta)^2$. We use $x=(\zeta,\bx)$ in the following of this paper.
\subsection{Hamiltonian Structure}
We use the standard definition of vierbein $g_{\mu\nu}=\eta_{ab}e^a_\mu e^\nu_b$. In the present paper,  $e$ sometimes denotes vierbein, sometimes denote $e=\sqrt{-g}$ and sometimes denote the base of natural exponential, depending on the contexts.
Denoting $\hat{\nabla}_a=e^\mu_a\nabla_\mu$ (here $e^\mu_a$ plays the role of parameter $\lambda(t)$ in \cite{Landau-Lifshitz1951}) and $\nabla_\mu$ is the standard covariant derivative, Lagrange of a free real Klein-Gordon field is 
\begin{align}
\mathscr{L}=&\frac{1}{2}(\hat{\nabla}^a\phi\hat{\nabla}_a\phi-\mKG^2\phi^2)\nonumber\\
=&\frac{1}{2}(\hat{\nabla}^0\phi\hat{\nabla}_0\phi+\hat{\nabla}^{a'}\phi\hat{\nabla}_{a'}\phi-\mKG^2\phi^2)
\end{align}
Defining \cite{Goldstein1980}
\beq
\Pi=\frac{\p\mathscr{L}}{\p(\hat{\nabla}_0\phi)}
\eeq
and assuming the existence of the inverse (for our case, it is apparent)
\beq
  \hat{\nabla}_0\phi= \hat{\nabla}_0\phi(\phi, \hat{\nabla}_i\phi,\Pi;x)
  \eeq
  , the Hamiltonian is defined in the standard way
\beq
\mathscr{H}=\hat{\nabla}_0 \phi \cdot\Pi-\mathscr{L}
\eeq
Thus
\beq
\frac{\p\mathscr{H}}{\p\Pi}=\hat{\nabla}_0 \phi +\Pi\frac{\p(\hat{\nabla}_0 \phi)}{\p\Pi}-\frac{\p\mathscr{L}}{\p(\hat{\nabla}_0 \phi)}\frac{\p(\hat{\nabla}_0 \phi)}{\p\Pi}=\hat{\nabla}_0 \phi 
\eeq
\beq
\frac{\p\mathscr{H}}{\p\phi}=\Pi\frac{\p(\hat{\nabla}_0 \phi)}{\p\phi}-\frac{\p\mathscr{L}}{\p(\hat{\nabla}_0 \phi)}\frac{\p(\hat{\nabla}_0 \phi)}{\p\phi}-\frac{\p\mathscr{L}}{\p\phi}=-\frac{\p\mathscr{L}}{\p\phi}
\eeq
Using the Euler-Lagrange eq., we have
\begin{eqnarray}
\frac{\p\mathscr{L}}{\p\phi}
&=&\nabla_\mu\left[e^\mu_0\frac{\p\mathscr{L}}{\p(\hat{\nabla}_0\phi)}\right]+
\nabla_\mu\left[e^\mu_{a'}\frac{\p\mathscr{L}}{\p(\hat{\nabla}_{a'}\phi)}\right]
\end{eqnarray}
(here the primes indices such as $a'$ run through 1,2,3.) we have
\begin{align}
\frac{\p\mathscr{H}}{\p\phi}=&-\nabla_\mu\frac{\p\mathscr{L}}{\p\nabla_\mu\phi}\nonumber\\
=&-\nabla_\mu\left[e^\mu_0\frac{\p\mathscr{L}}{\p(\hat{\nabla}_0\phi)}\right]-
\nabla_\mu\left[e^\mu_{a'}\frac{\p\mathscr{L}}{\p(\hat{\nabla}_{a'}\phi)}\right]\nonumber\\
=&-\nabla_\mu\left[e^\mu_0\Pi\right]-
\nabla_\mu\left[e^\mu_{a'}\frac{\p\mathscr{L}}{\p(\hat{\nabla}_{a'}\phi)}\right]
\end{align}
Since
\begin{align}
\frac{\p\mathscr{H}}{\p(\hat{\nabla}_{a'}\phi)}=&\Pi\frac{\p(\hat{\nabla}_0\phi)}{\p(\hat{\nabla}_{a'}\phi)}-\frac{\p\mathscr{L}}{\p(\hat{\nabla}_0\phi)}
\frac{\p(\hat{\nabla}_0\phi)}{\p(\hat{\nabla}_{a'}\phi)}-\frac{\p\mathscr{L}}{\p(\hat{\nabla}_{a'}\phi)}\nonumber\\
=&-\frac{\p\mathscr{L}}{\p(\hat{\nabla}_{a'}\phi)}
\end{align}
we find
\beq
\nabla_\mu\left[e^\mu_0\Pi\right]=-\frac{\p\mathscr{H}}{\p\phi}+\nabla_\mu\left[e^\mu_{a'}\frac{\p\mathscr{H}}{\p(\hat{\nabla}_{a'}\phi)}\right]
\eeq
Introducing {\it functional derivative} 
\beq
\frac{\delta}{\delta\psi}=\frac{\p}{\p\psi}-\nabla_\mu\left[e^\mu_{a'}\frac{\p}{\p(\hat{\nabla}_{a'}\psi)}\right]
\eeq
we have
\beq
\hat{\nabla}_0\phi=\frac{\delta \mathscr{H}}{\delta\Pi}
\eeq
and
\beq
\nabla_\mu\left[e^\mu_0\Pi\right]=-\frac{\delta\mathscr{H}}{\delta\phi};
\eeq
These can be cast into the conventional formalism. Defining Cauchy surface $\Sigma: f(x)=\zeta=$const, we have functional
\beq
H[\phi,\Pi;\zeta]:=\int_\Sigma  d\sigma\mathscr{H}
\eeq
where
\begin{align}
\mathscr{H}=&\frac{1}{2}(\Pi^2-\hat{\nabla}^{a'}\phi\hat{\nabla}_{a'}\phi+\mKG^2\phi^2)\\
d\sigma_{|\Sigma}:=&d\sigma_\mu n^\mu=\frac{1}{3!}e\vep_{\mu\nu\alpha\beta}dx^\nu\wedge dx^\alpha\wedge dx^\beta n^\mu\nonumber\\
=&C^{3/2}(\zeta)d^3\bx
\end{align}
here $n^\mu=C^{-1/2}(\zeta)(1,0,0,0)$ is the unit normal to $\Sigma$.  For two functionals $F[\phi,\Pi;\zeta]=\int_\Sigma  d\sigma\mathscr{F}, G[\phi,\Pi;\zeta]=\int_\Sigma  d\sigma\mathscr{G}$, the Poisson bracket is defined as
\beq
\left\{ F,G\right\}^{\text{P.B.}}:=\int_\Sigma d\sigma \left(\frac{\delta \mathscr{F}}{\delta\phi(x)}\frac{\delta \mathscr{G}}{\delta\Pi(x)}-\frac{\delta \mathscr{G}}{\delta\phi(x)}\frac{\delta \mathscr{F}}{\delta\Pi(x)}\right)
\eeq   
Since $\phi(x)=\int_{\Sigma} d\sigma' \delta^3(x-x')C^{-3/2}(\zeta)\phi(x'), \Pi(y)=\int_{\Sigma} d\sigma' \delta^3(y-y')C^{-3/2}(\zeta)\Pi(y')$, we have then
\beq
\left\{\phi(x),\Pi(y)\right\}^{\text{P.B.}}=\delta^3(Z^i)=\frac{1}{e'}\delta^3(\bx-\by)
\eeq
Here $Z^i$ are the local flat coordinates defined as $Z^i=e^i_\mu (x-y)^\mu$  for $y$ very close to $x$ and $e'=\det(e^{a'}_{\mu'})$. 
Hence we have
\begin{align}
\hat{\nabla}_0\phi& =  \left\{\phi(x), H\right\}_{x^0=\zeta}^{\text{P.B.}} \\
\hat{\nabla}_0\Pi&=\left\{\Pi(x), H\right\}_{x^0=\zeta}^{\text{P.B.}}-\omega_0\Pi
\end{align}
where
\beq
\omega_0=\nabla_\mu e^\mu_0=\frac{1}{C^2(\zeta)}\p_\zeta (C^2(\zeta) C^{-1/2}(\zeta))
\eeq
\subsection{Appropriate Phase Space Variables}
Since the equation for $\Pi$ involves a term $\omega_0\Pi$, which does not exist in conventional canonical equations of motion, we seek new definitions of canonical variables. Defining $\pi:=e^\Omega \Pi$
where
$
\hat{\nabla}_0\Omega=\omega_0
$
which is a scalar, we have
$
\Omega=\ln C^{3/2}(\zeta)
$, 
then
\begin{align}
e^\mu_0\nabla_\mu\pi=&e^\Omega \hat{\nabla}_0\Pi+\hat{\nabla}_0e^\Omega e^\Omega\Pi=
e^\Omega\hat{\nabla}_0\Pi+\hat{\nabla}_0\Omega\cdot \pi\nonumber\\
=&e^\Omega\left(\left\{\Pi(x), H\right\}_{x^0=\zeta}^{\text{P.B.}}-\omega_0\Pi\right)+\omega_0\cdot \pi\nonumber\\
=&\left\{\pi(x), H\right\}_{x^0=\zeta}^{\text{P.B.}}
\end{align}
In terms of $\phi, \pi$, we have
\begin{align}
\hat{\nabla}_0\phi& =  \left\{\phi(x), H[\phi,\pi;\zeta]\right\}_{x^0=\zeta}^{\text{P.B.}} \\
\hat{\nabla}_0\pi&=\left\{\pi(x), H[\phi,\pi;\zeta]\right\}_{x^0=\zeta}^{\text{P.B.}} \label{Heom}
\end{align}
\beq
\left\{\phi(x),\pi(y)\right\}^{\text{P.B.}}=e^\Omega\delta^3(Z^i)=\delta^3(\bx-\by) \label{PB}
\eeq
It is worthwhile to point out the two benefits  of using $\pi$ instead of $\Pi$. First,  the classical equation of $\pi$ does not have the term $\omega_0\Pi$ on the right. Second the, the Poisson bracket eq.(\ref{PB}) does not dependent on time, a nice feature when we transform from Heisenberg picture to Schr\"{o}dinger picture later on.  Since in Schr\"{o}dinger picture, canonical conjugate fields do not depend on time and hence should not do their commutators . 

 In general, for any $O[\phi,\pi;\lambda(x)]$
\beq
O[\phi,\pi;\lambda(x)]=\int_\Sigma d\sigma  \mathscr{O}(\phi,\pi;\lambda(x))
\eeq
which is a function of $\zeta$ and a functional of $\phi,\pi$
\begin{align}
 \hat{d}_0O[\phi,\pi;\lambda(x)](\zeta)& =\int_\Sigma e^\zeta_0(x)\p_\zeta(d\sigma)\mathscr{O}+\left\{O, H\right\}\nonumber\\
 &+\int_\Sigma d\sigma\left(- \frac{\delta  \mathscr{O}}{\delta\pi(x)} \omega_0\pi+\frac{\delta  \mathscr{O}}{\delta\lambda(x)}\hat{\nabla}_0\lambda\right)
  \end{align}
 where $\hat{d}_0$ is defined as $e^\zeta_0(\zeta)d/d\zeta$, bearing mind that $\zeta=\text{const.}$ defines the surface $\Sigma$, i.e., the l.h.s. depends on the surface $\Sigma$. In particular,
\begin{align}
 \hat{d}_0H[\phi,\pi;\cdots]=&\int_\Sigma e^\zeta_0(x)\p_\zeta(d\sigma)\mathscr{H}\nonumber\\
 &-\int_\Sigma d\sigma \left( \frac{\delta  \mathscr{H}}{\delta\pi(x)} \omega_0\pi+\frac{\delta  \mathscr{H}}{\delta e^\mu_a}\hat{\nabla}_0e^\mu_a\right)
\end{align}
In terms of canonical variables $\phi,\pi$
\begin{align}
\mathscr{H}=&\frac{1}{2}(e^{-2\Omega}\pi^2-\hat{\nabla}^{a'}\phi\hat{\nabla}_{a'}\phi+\mKG^2\phi^2)
\end{align}
Defining new Poisson bracket
\beq
\left\{ F,G\right\}^\text{new}:=\int_\Sigma d\sigma   e^{\Omega(x)}\left(\frac{\delta \mathscr{F}}{\delta\phi(x)}\frac{\delta \mathscr{G}}{\delta\pi(x)}-\frac{\delta \mathscr{G}}{\delta\phi(x)}\frac{\delta \mathscr{F}}{\delta\pi(x)}\right)
\eeq   
then
\begin{align}
\hat{\nabla}_0\phi& = \left\{\phi(x), H[\phi,\pi;\zeta]\right\}_{x^0=\zeta}^{\text{new}} \\
\hat{\nabla}_0\pi&= \left\{\tilde{\Pi}(x), H[\phi,\pi;\zeta]\right\}_{x^0=\zeta} ^{\text{new}}
\end{align}
 \section{Schr\"{o}dinger Picture}
 Upon quantization, the classical canonical variables $\phi,\pi$ are replaced by operators $\hat{\phi},\hat{\pi}$  in a Hilbert space and Poisson brackets become commutators.
   In standard quantized Klein-Gordon field theory in Minkowski spacetime, the Hamiltonian is time-independent and three pictures can be utilized. Similarly, 
we can define Schr\"{o}dinger picture $O=\phi,\pi$
\begin{align}
O^{\text{S}}(\zeta)=&\hat{T}^{-1}e^{i\int^\zeta_\ell H(\eta)e^0_\zeta(\eta)d\eta}O (\zeta)\hat{T}e^{-i\int^\zeta_\ell H(\eta)e^0_\zeta(\eta)d\eta}
\end{align}
where $\hat{T}$ is the time-ordering operator define as  $\hat{T}\phi(\zeta_1)\phi(\zeta_2)=\phi(\zeta_1)\phi(\zeta_2)\theta(\zeta_2-\zeta_1)+\phi(\zeta_2)\phi(\zeta_1)\theta(\zeta_1-\zeta_2).$  The two pictures agree at $\zeta=\ell$
\beq
\phi^{\text{S}}(\bx)=\phi(\ell,\bx)
\eeq
For Hamiltonian
\begin{align}
H^{\text{S}}(\zeta)=&\hat{T}^{-1}e^{i\int^\zeta_\ell H(\eta)e^0_\zeta(\eta)d\eta}H (\zeta)\hat{T}e^{-i\int^\zeta_\ell H(\eta)e^0_\zeta(\eta)d\eta}\nonumber\\
=&H[\phi^{\text{S}}(\bx),\pi^{\text{S}}(\bx);e^\mu_a(x),\Omega(x)](\zeta)
\end{align}
and $
H^{\text{S}}(\ell)=H(\ell)
$. 
Since
\beq
i\hat{\nabla}_0\phi^{\text{S}}(x)=0
\eeq  
So $\phi^{\text{S}}, \pi^{\text{S}}$ are time-independent, playing the roles of $x,p$  while $\phi,\pi$ play the role $x_\pm,p_\pm$ in \cite{Landovitz1979} :
\begin{align}
H(t)=&f(t)p^2/2m+g(t)\frac{1}{2}m\omega^2_0x^2,\nonumber\\
H_+(t)=&f(t)p_+^2(t)/2m+g(t)\frac{1}{2}m\omega^2_0x_+^2(t)\nonumber
\end{align}
\begin{align}
dH(t)/dt=&\dot{f}(t)p^2/2m+\dot{g}(t)\frac{1}{2}m\omega^2_0x^2,\nonumber\\
dH_+(t)/dt=&\dot{f}(t)p_+^2(t)/2m+\dot{g}(t)\frac{1}{2}m\omega^2_0x_+^2(t)\nonumber
\end{align}
Though the initial condition $H(0)=H_+(0), H(t)\not=H_+(t)$ since $p_+(t), x_+(t)$ depend on time $t$. The time-dependence of $H^{\text{S}}$ is
\begin{align}
i\hat{d}_0H^{\text{S}}(\zeta)
=ie^\zeta_0\frac{\p}{\p\zeta} H[\phi^{\text{S}}(\bx),\pi^{\text{S}}(\bx);e^\mu_a(x),\Omega(x)](\zeta)
\end{align}
Quantization is carried out by the correspondence
\begin{align}
[\hat{\phi}(\bx), \hat{\pi}(\by)]_{x^0=y^0}:=&i\hbar  \left\{\phi(x), \pi(y)\right\}_{x^0=y^0}^{\text{new}}=i\hbar\delta^3(\bx-\by)
\end{align}
Quantum mechanical Schr\"odinger  eq. for  wave-functional
\beq
i\hat{\nabla}_0\Psi[\phi(\bx),\zeta]=H^{\text{S}}[\phi(\bx),\pi(\bx);\zeta]\Psi[\phi(\bx),\zeta]
\eeq
where in $H^{\text{S}}$  (As in conventional quantum field theories, this is not unique!)
$
\pi(\bx)\mapsto -i\hbar \delta/\delta\phi(\bx).
$
So 
\begin{align}
&H^{\text{S}}(\zeta)\nonumber\\
=&\frac{1}{2}\int_\Sigma d\sigma \left[ e^{-2\Omega}\pi^2(\bx)-\hat{\nabla}^{a'}\phi(\bx)\hat{\nabla}_{a'}\phi(\bx) +\mKG^2\phi^2(\bx))\right]\nonumber\\
=&\frac{1}{2}\int _\Sigma d\sigma \Big[ C^{-3}(\zeta)\left(-i\hbar\frac{\delta}{\delta \phi(\bx)}\right)^2-\hat{\nabla}^{a'}\phi(\bx)\hat{\nabla}_{a'}\phi(\bx) \nonumber\\
&+\mKG^2\phi^2(\bx)\Big] \label{SchHa}
\end{align}
This is slightly different from \cite{Jackiw}. The Gaussian-type Schr\"{o}dinger Wave Functional
$
\Psi[\phi(\bx),\zeta]=\eta e^{-G[\phi(\bx),\zeta]}
$
satisfies by eq.(\ref{SchHa})
\begin{align}
&H^{\text{S}}[\phi(\bx),\pi(\bx);\zeta]\Psi[\phi(\bx),\zeta]\nonumber\\
=&\frac{1}{2}\int d\sigma \Big[  C^{-3}(\zeta)\left(\frac{\delta^2G}{\delta\phi(\bx)^2}-(\frac{\delta G}{\delta\phi(\bx)})^2\right)\nonumber\\
&-\hat{\nabla}^{a'}\phi\hat{\nabla}_{a'}\phi +\mKG^2\phi^2\Big]\Psi[\phi(\bx),\zeta]
\end{align}
Assuming
\beq
G[\phi(\bx),\zeta]=iE(\zeta)+\int d^3\bx\int d^3\by \phi(\bx)f(\bx,\by;\zeta)\phi(\by)
\eeq
then
\begin{align}
&-2C^{-3}(\zeta)\int d\sigma d^3\by d^3\bz \phi(\bz)f(\bz,\bx;\zeta)f(\bx,\by;\zeta)\phi(\by)\nonumber\\
&+\frac{1}{2}\int d\sigma \phi(\bx)\left[  e^\mu_{a'}e^\nu_{b'}\eta^{a'b'}\nabla_\mu\nabla_\nu +\mKG^2\right]\phi(\bx)\nonumber\\
=&i e^\zeta_0(\zeta)
\int d^3\bx\int d^3\by \phi(\bx)\p_\zeta  f(\bx,\by;\zeta)\phi(\by)
\end{align}
Using Fourier transformation
$
f(\bx,\by;\zeta)=\sum_{\bk}f_{\bk}(\zeta) e^{i\bk\cdot (\bx-\by)}
$
we arrive
\begin{align}
-2C^{-1}(\zeta)(2\pi)^3  f^2_{\bk}(\zeta)+\frac{1}{2}C(\zeta)(2\pi)^3\omega_{\bk}^2(\zeta)=i \p_\zeta  f_{\bk}(\zeta) 
\end{align}
which is a Riccati equation
and can be transformed into a  Bessel-type equation.

\section{Fundamental Solutions for Free Field}
\subsection{Basis Solutions}
As has been long since known, the 
 equation $(\nabla^2+\mKG^2)\phi=0$ in terms of co-moving coordinates of de Sitter spacetime reads \cite{Birrell}
\beq
C^{-2}(\zeta)\p_\zeta(C(\zeta)\p_\zeta\phi)+\p_i(C^{-1}(\zeta)\eta^{ij}\p_j\phi)+\mKG^2\phi=0
\eeq
and the solutions are
$
\Phi_{\uk}={\mathscr{Y}}_{\uk}(\boldsymbol{x})C^{-1/2}(\zeta)f_{\uk}(\zeta)$  where ${\mathscr{Y}}_{\uk}(\boldsymbol{x})=(2\pi)^{-3/2}e^{i\boldsymbol{k}\cdot\boldsymbol{ x}},\,\,\,\,\,\,\,
\underline{k}=\boldsymbol{k}=(k_1,k_2,k_3), \,\,\,\,(-\infty<k_i<\infty) $.  
$f_{\uk}(\zeta)$ satisfies
\beq
\ddot{f}_{\uk}(\zeta)+\Big[
\bk^2+\frac{1}{4}(\frac{\dot{C}}{C})^2-\frac{1}{2}\frac{\ddot{C}}{C}
+m^2_{\text{KG}}C(\zeta)\Big]f_{\uk}(\zeta)
=0
\eeq
For $C(\zeta)=A\zeta^w$, 
\beq
\ddot{f}_{\uk}(\zeta)+\Big[
\bk^2+\frac{1}{4}\frac{w^2}{\zeta^2}-\frac{1}{2}\frac{w(w-1)}{\zeta^2}
+m^2_{\text{KG}}A\zeta^w\Big]f_{\uk}(\zeta)
=0
\eeq
In our case $w=-2$, we have
\beq
\ddot{f}_{\uk}(\zeta)+(\bk^2-\frac{2-\ell^2 \mKG^2}{\zeta^2})f_{\uk}(\zeta)=0 \label{dSf}
\eeq
Let
$
f_{\uk}(\zeta)=\zeta^{1/2}J(\zeta), 
$
then (where $z=|\bk|\zeta$)
\beq
\ddot{J}(z)+\frac{1}{z}\dot{J}(z)+(1-\frac{\nu^2}{z^2})J=0
\eeq
with
\beq
\nu^2=\frac{9}{4}-\ell^2m^2_{\text{KG}}
\eeq
(For electron, the Compton wave length $\lambda_e=\hbar/m_ec=3.86\times 10^{-13} \text{m}$. So $\ell/\lambda_e\sim 10^{40}\gg 1$.
)
Therefore, for $9/4>\ell^2m^2_{\text{KG}}$, (i.e., the {\it complementary series})  we have for $\nu\not= n$,  (denoting $k=|\bk|$)
\beq
f_{\uk}=\zeta^{1/2}\Big[\alpha_{1\uk}H^{(1)}_\nu(k\zeta)+\alpha_{2\uk}H^{(2)}_\nu(k\zeta)
\Big]
\eeq
For integer $\nu$,
\beq
f_{\uk}=\zeta^{1/2}\Big[\alpha_{1\uk}J_n(k\zeta)+\alpha_{2\uk}Y_n(k\zeta)
\Big]
\eeq
Hence for non-integer $\nu$ we have
\beq
\Phi_{\uk}\sim(2\pi)^{-3/2}e^{i\boldsymbol{k}\cdot\boldsymbol{ x}}\frac{\zeta^{3/2}}{\ell}\Big[\alpha_{1\uk}H^{(1)}_\nu(k\zeta)+\alpha_{2\uk}H^{(2)}_\nu(k\zeta)
\Big]
\eeq
As we choose $e^{i(-\omega t+\bk\cdot\bx)}$ in Minkowski space-time, we write 
$
\Phi_{\bk}(\zeta,\bx)
=:g_{\bk}(\zeta)e^{i\bk\cdot\bx},
$
where
\beq
g_{\bk}(\zeta)
=(2\pi)^{-3/2}A(\bk) \zeta^{3/2}H^{(2)}_\nu(k\zeta)
\eeq
where $A(\bk)$ will be determined later by the normalization of Klein-Gordon product. For $\zeta\rightarrow 0$
\beq
g_{\bk}(\zeta)=(2\pi)^{-3/2}A(\bk) \zeta^{3/2}H^{(2)}_\nu(k\zeta)\rightarrow \zeta^{3/2-\nu}\sim 0
\eeq
which is reminiscent of $\lim_{t\rightarrow \infty}e^{i\omega t}\sim 0$ in the sense Riemann-Lebesgue Lemma of  generalized functions.
\indent For $9/4<\ell^2m^2_{\text{KG}}$, (i.e., the {\it principal series}),  let $\nu=i\mu$, we have
\beq
f_{\uk}=\zeta^{1/2}\Big[\beta_{1\uk}H^{(1)}_{i\mu}(k\zeta)+\beta_{2\uk}H^{(2)}_{i\mu}(k\zeta)
\Big]
\eeq
 In this work, we consider the complementary series only and zero-mass cases can be included.
\subsection{Klein-Gordon Current and Mode Expansion}
Conventionally, the Klein-Gordon current defined as \cite{Crispino2007}
\beq
J^\mu_{(f_A,f_B)}:=f^*_A(x)\nabla^\mu f_B(x)-f_B(x)\nabla^\mu f^*_A(x)
\eeq
is conserved.
\beq
\nabla_\mu J^\mu_{(f_A,f_B)}=0
\eeq
Hence the quantity
\beq
( f_A|f_B)_{\rm KG}:=i\int_\Sigma d\sigma_\mu \,\,J^\mu_{(f_A,f_B)}=i\int_\Sigma d\sigma f^*_A(x)\overleftrightarrow{\mathpzc{L}_n}f_B(x)
\eeq
is constant over the foliations of surfaces $\Sigma$. We have
\beq
( f_A|f_B)^*_{\rm KG}=( f_B|f_A)_{\rm KG}
\eeq
Suppose with coordinates $x$, there are a complete set of solutions $f_i,f^*_i$to the K-G equation satisfying
\begin{align}
( f_i|f_j)_{\rm KG}=&-( f^*_i|f^*_j)_{\rm KG}=\delta_{ij}\\
( f_i^*|f_j)_{\rm KG}=&( f_i| f^*_j)_{\rm KG}=0
\end{align}
and 
\beq
\sum_i[f_i(x)f_i^*(y)-f^*_i(x)f_i(y)]=0
\eeq
 Completeness requires that any K-G solution $g(x)$ can be expanded as
\beq
g(x)=\sum_i (g_if_i+g^*_if^*_i)
\eeq
where
\beq
g_i=(f_i|g)_{\rm KG}, \,\,\,\,\,\,\,\,\,\,\,\,
g^*_i=-(f^*_i|g)_{\rm KG}
\eeq
So we have resolution of identity
\beq
\sum_i  |f_i)( f_i|-\sum_i |f^*_i)( f^*_i|=1
\eeq
or
\begin{align}
&i\sum_i \big[f_i(x)  (f^*_i(y)n^\mu(y)\nabla_\mu-n^\mu(y)\nabla_\mu f^*_i(y))\nonumber\\
&-f^*_i(x)(f_i(y)n^\mu(y)\nabla_\mu-n^\mu(y)\nabla_\mu f_i(y))]_{|\Sigma}=\delta^3(Z^i)
\end{align}
Since if $f_i (x)$ is a solution, so must be $f ^*_i(x)$, we have therefore
or
\begin{align}
&\sum_i \big[f_i(x)  n^\mu(y)\nabla_\mu f^*_i(y)-f^*_i(x)n^\mu(y)\nabla_\mu f_i(y)]_{|\Sigma}\nonumber\\
&=i\delta^3(Z^i)
\end{align}
For free field, we can expand
\beq
\phi(x)=\sum_i\big[a_if_i+a^\dag_if^*_i\big]
\eeq
then
\beq
a_i=( f_i|\phi)_{\rm KG},\,\,\,\,\,\,\,\,\,
a^\dag_j=-( f^*_j|\phi)_{\rm KG}
\eeq
Since
\begin{align}
( f_i|\phi)_{\rm KG}=&i\int_{\Sigma} d\sigma n_\mu (f^*_i\nabla^\mu\phi-\phi\nabla^\mu f^*_i)\nonumber\\
=&i\int_{\Sigma} d\sigma (f^*_i\Pi-n_\mu\phi\nabla^\mu f^*_i)\\
( f^*_j|\phi)_{\rm KG}=&i\int_{\Sigma} d\sigma n_\mu (f_j\nabla^\mu\phi-\phi\nabla^\mu f_j)\nonumber\\
=&i\int_{\Sigma} d\sigma (f_j\Pi-n_\mu\phi\nabla^\mu f_j)
\end{align}
Hence
\beq
[a_i,a^\dag_j]=i\int_{\Sigma_{\text{KG}}} d\sigma f^*_i\overleftrightarrow{\mathpzc{L}_n} f_j=(f_i|f_j)_{\rm KG}=\delta_{ij}
\eeq
\beq
\Pi(y)=n^\mu\sum_j\nabla_\mu (a_jf_j+a^\dag_jf^*_j)
\eeq
\begin{align}
&[\phi(x),\Pi(y)]_{|\Sigma}\nonumber\\
=&\sum_i \big[f_i(x)\nabla_\mu f^*_i(y)-f^*_i(x)\nabla_\mu f_i(y)\big]n^\mu(y)
=i\delta^3(Z^i)
\end{align}

The K-G product on $\Sigma$ defined by $\zeta=\zeta_0$  is
\begin{align}
&(\Phi_{\bk}|\Phi_{\bk'})_{\rm KG}\nonumber\\
=&i\int_{\zeta=\zeta_0}d\sigma_\mu (\Phi^*_{\bk} \nabla^\mu\Phi_{\bk'}    -
 \nabla^\mu\Phi^*_{\bk} \Phi_{\bk'} )\nonumber\\
 =&i\int \delta(\zeta-\zeta_0)  (\Phi^*_{\bk} \nabla_\zeta\Phi_{\bk'}    -
 \nabla_\zeta\Phi^*_{\bk} \Phi_{\bk'} )  C(\zeta)d\zeta d^3\bx\nonumber\\
   \end{align}
 Using the expression of  $\Phi_{\bk} $, we have
 \begin{align}
&(\Phi_{\bk}|\Phi_{\bk'})_{\rm KG}\nonumber\\
 =&i|A(\bk)|^2\delta^3(\bk-\bk')\ell^2\zeta_0 \Big[H^{(2)*}_\nu(k\zeta_0) \p_{\zeta_0}\big[H^{(2)}_\nu(k\zeta_0) \big] \nonumber\\
 & -
 \p_{\zeta_0}\big[_0H^{(2)*}_\nu(k\zeta_0)\big]H^{(2)}_\nu(k\zeta_0) \Big] \nonumber\\
    \end{align}  
 For real or imaginary $\nu$, the Wronskian of any two solutions $Z_\nu^1, Z_\nu^2$ to Bessel equation  satisfies
 \beq
 zW[Z^{1*}_\nu, Z^2_\nu]=\text{Const.}
 \eeq
  So we must have
  \beq
  z \Big[H^{(2)*}_\nu(z) \p_z\big[H^{(2)}_\nu(z) \big]  -
 \p_z\big[H^{(2)*}_\nu(z)\big]H^{(2)}_\nu(z) \Big] =\text{Const.}
   \eeq

 To find the constant, we can use the asymptotic behavior, 
 $H^{(2)}_\nu(z)\sim\sqrt{2/\pi z}e^{-i(z-\frac{\nu\pi}{2}-\frac{\pi}{4})}
 $
 and take limit for large $z$, one obtains
 \begin{align}
  &z \Big[H^{(2)*}_\nu(z) \p_z\big[H^{(2)}_\nu(z) \big]  -
 \p_z\big[H^{(2)*}_\nu(z)\big]H^{(2)}_\nu(z) \Big] \nonumber\\
 =&\frac{-4i}{\pi}e^{\frac{\pi i}{2}(\nu-\nu^*)} \label{HankelWronski}
   \end{align}
 hence
 \beq
 (\Phi_{\bk}|\Phi_{\bk'})_{\rm KG}=|A(\bk)|^2\delta^3(\bk-\bk')\ell^2\frac{4}{\pi}e^{\frac{\pi i}{2}(\nu-\nu^*)};  \eeq
which is apparently $\zeta_0$-independent.  Therefore we should choose
\beq
 A(\bk)=\frac{\sqrt{\pi}}{2\ell}e^{-\frac{\pi i}{4}(\nu-\nu^*)}.
  \eeq

\section{Second Quantization: Free Field}
   Here we use discrete notation for mode expansion.  
\begin{align}
\phi(x)=&\sum_{\bk}(a_{\bk}\Phi_{\bk}+a^\dag_{\bk}\Phi^*_{\bk})\\
\Pi=&e^\mu_0\nabla_\mu\phi
=C^{-1/2}(\zeta)\sum_{\bp}(a_{\bp}\nabla_\zeta\Phi_{\bp}+a^\dag_{\bp}\nabla_\zeta\Phi^*_{\bp})
\end{align}
we can calculate the basic commutator
\begin{align}
&\left[\phi(x),\Pi(y)\right]\nonumber\\
&=C^{-1/2}(\zeta)\sum_{\bk}\left(\Phi_{\bk}(x)\nabla_\zeta \Phi^*_{\bk}(y)-\Phi^*_{\bk}(x)\nabla_\zeta \Phi_{\bk}(y)\right)\nonumber\\
&=(2\pi)^{-3/2}C^{-1/2}(\zeta)\frac{3}{2}\zeta^{1/2}\sum_{\bk}\Big[\Phi_{-\bk}(x)A^*(-\bk)e^{i\bk\cdot\by}H^{(2)*}_\nu(k\zeta))\nonumber\\
&-\Phi^*_{\bk}(x)A(\bk)e^{i\bk\cdot\by}H^{(2)}_\nu(k\zeta))\Big]\nonumber\\
&+(2\pi)^{-3/2}C^{-1/2}(\zeta)\zeta^{3/2}\sum_{\bk}\Big[\Phi_{-\bk}(x)A^*(-\bk)e^{i\bk\cdot\by}\nabla_\zeta H^{(2)*}_\nu(k\zeta))
\nonumber\\
&-\Phi^*_{\bk}(x)A(\bk)e^{i\bk\cdot\by}\nabla_\zeta (H^{(2)}_\nu(k\zeta))\Big]
\end{align}
The first term vanishes, hence
\begin{align}
&\left[\phi(x),\Pi(y)\right]=(2\pi)^{-3}C^{-1/2}(\zeta)\zeta^{3}\sum_{\bk}|A(\bk)|^2\nonumber\\
&\times \Big[e^{i\bk\cdot(\by-\bx)}H^{(2)}_\nu(k\zeta )\nabla_\zeta H^{(2)*}_\nu(k\zeta))\nonumber\\
&-e^{i\bk\cdot(\by-\bx)}H^{(2)*}_\nu(k\zeta)\nabla_\zeta (H^{(2)}_\nu(k\zeta))\Big]\nonumber\\
\end{align}
Use eq.(\ref{HankelWronski})
\begin{eqnarray}
\left[\phi(x),\Pi(y)\right]_{|\Sigma}
&=&iC^{-3/2}(\zeta)\delta^3(\by-\bx)
\end{eqnarray}
Similarly
\begin{align}
&\left[\Pi(x),\Pi(y)\right]\nonumber\\
=&C^{-1}(\zeta)\sum_{\bk,\bp}\Big[a_{\bp}\nabla_\zeta\Phi_{\bp}(x)+a^\dag_{\bp}\nabla_\zeta\Phi^*_{\bp}(x),  a_{\bk}\nabla_\zeta\Phi_{\bk}(y)\nonumber\\ &+a^\dag_{\bk}\nabla_\zeta\Phi^*_{\bk}(y)\Big]\nonumber\\
=&C^{-1}(\zeta)\sum_{\bp}\left[\dot{g}_p(\zeta)\dot{g}^*_p(\zeta)e^{i\bp\cdot(\bx-\by)}-\dot{g}^*_p(\zeta)\dot{g}_p(\zeta)e^{-i\bp\cdot(\bx-\by)}\right]=0
\end{align}
\subsection{Heisenberg 2nd Quantized Hamiltonian}
As in time-dependent harmonic oscillators, $H\not= H_\pm$\cite{Landovitz1979}, we need to discuss Hamiltonian in Heisenberg and Schr\"{o}dinger picture separately.
Denoting $\p_\zeta g_{\bk}(\zeta)=\dot{g}_{\bk}(\zeta)$   
\begin{align}
H&=\frac{1}{2}\int C^{3/2}(\zeta)d^3\bx [C^{-1}(\zeta)\sum_{\bk,\bp}(a_{\bp}\dot{g}_{\bp}e^{i\bp\cdot\bx}+a^\dag_{\bp}\dot{g}^*_{\bp}e^{-i\bp\cdot\bx})
\nonumber\\
&\times (a_{\bk}\dot{g}_{\bk}e^{i\bk\cdot\bx}+a^\dag_{\bk}\dot{g}^*_{\bk}e^{-i\bk\cdot\bx})\nonumber\\
&+C^{-1}(\zeta)\sum_{\bk,\bp}(i\bp)\cdot(i\bk)(a_{\bp}g_{\bp}e^{i\bp\cdot\bx}-a^\dag_{\bp}g^*_{\bp}e^{-i\bp\cdot\bx})\nonumber\\
&\times(a_{\bk}g_{\bk}e^{i\bk\cdot\bx}-a^\dag_{\bk}g^*_{\bk}e^{-i\bk\cdot\bx})\nonumber\\
&+\mKG^2\sum_{\bk,\bp}(a_{\bp}g_{\bp}e^{i\bp\cdot\bx}+a^\dag_{\bp}g^*_{\bp}e^{-i\bp\cdot\bx})
(a_{\bk}g_{\bk}e^{i\bk\cdot\bx}+a^\dag_{\bk}g^*_{\bk}e^{-i\bk\cdot\bx})]
\end{align}   
Writing
  \begin{align}
H(\zeta)=&\frac{1}{2}\sum_{\bk}\vep_{\bk}(\zeta)(a^\dag_{\bk} a_{\bk}+a_{\bk}a^\dag_{\bk})\nonumber\\
&+\frac{1}{2}\sum_{\bk}\left[\Delta_{\bk}(\zeta)a_{\bk}a_{-\bk}+\Delta^*_{\bk}(\zeta)a^\dag_{-\bk}a^\dag_{\bk}\right]
  \end{align}
where
\begin{align}
\vep_{\bk}&=(2\pi)^3C^{1/2}(\zeta)\left[|\dot{g}_{\bk}|^2+\left(\bk^2+C(\zeta)\mKG^2\right)|g_{\bk}|^2\right]\\
\Delta_{\bk}&=(2\pi)^3C^{1/2}(\zeta)\left[\dot{g}_{\bk}^2+\left(\bk^2+C(\zeta)\mKG^2\right)g_{\bk}^2\right]
\end{align}   
Our expression is slightly different from \cite{Akhmedov2014}. The Hamiltonian $H(\zeta)$ can be diagonalized by the Bogliubov transformation.
\beq
a_{\bk}=u^*_{-\bk}b_{\bk}-v_{\bk}b^\dag_{-\bk}, \quad a^\dag_{\bk}=u_{-\bk}b^\dag_{\bk}-v^*_{\bk}b_{-\bk}
\eeq
where
\beq
u_{\bk}=\sqrt{\frac{\vep_{\bk}+\omega_{\bk}(\zeta)}{2\omega_{\bk}(\zeta)}} ,  \quad v_{\bk}=\frac{\Delta^*_{\bk}}{\vep_{\bk}+\omega_{\bk}(\zeta)}u_{\bk}
\eeq
and $\omega_{\bk}(\zeta)=\sqrt{\vep_{\bk}^2-|\Delta_{\bk}|^2}$.  We have
\beq
H(\zeta)=\frac{1}{2}\sum_{\bk}\omega_{\bk}(\zeta)\left[b^\dag_{\bk}(\zeta) b_{\bk}(\zeta)+b_{\bk}(\zeta)b^\dag_{\bk}(\zeta)\right]
\eeq
According to standard quantum theory of many-body systems\cite{Fetter1971}, $b^\dag_{\bk}(\zeta)$ generate observed quasi-particles/excitations. 
From the inverse
\begin{align}
b_{\bk}(\zeta)=&u_{\bk}(\zeta) a_{\bk}+v_{\bk}(\zeta) a_{-\bk}^{\dag},\\
 \quad b_{\bk}^{\dag}(\zeta)=&u_{\bk}^{*}(\zeta) a_{\bk}^{\dag}+v_{\bk}^{*}(\zeta) a_{-\bk},
\end{align}
we have the commutation relations
\begin{align}
[b_{\bk}(\zeta_1),b_{\bp}^\dag(\zeta_2)]=&(u_{\bk}(\zeta_1)u^*_{\bk}(\zeta_2)-v_{\bk}(\zeta_1)v^*_{\bk}(\zeta_2))\delta^3(\bk-\bp)\\
[b_{\bk}(\zeta_1),b_{\bp}(\zeta_2)]=&(u_{\bk}(\zeta_1)v_{\bp}(\zeta_2)-v_{\bk}(\zeta_1)u_{\bp}(\zeta_2))\delta^3(\bk+\bp)
\end{align}
So $b_{\bk}(\zeta_1),b_{\bk}(\zeta_2)$ do not commute if $\zeta_1\not=\zeta_2$.
The momentum operata
\begin{align}
\bP&=\sum_{\bk}\bk a_{\bk}^\dag a_{\bk}=\sum_{\bk}\bk b^\dag_{\bk}b_{\bk}
\end{align}
$C^{-1/2}(\zeta)\bk$ is the measured momentum.  Again using eq.(\ref{HankelWronski})
\begin{align}
\dot{g}_{\bk}^* g_{\bk}-\dot{g}_{\bk}g^*_{\bk}=&(2\pi)^{-3}\ell^{-2}\zeta^2(-i)
\end{align}
and the expression 
\begin{align}
\omega_{\bk}^2=&-(2\pi)^6C(\zeta)\left(\bk^2+C(\zeta)\mKG^2\right)(\dot{g}_{\bk}^* g_{\bk}-\dot{g}_{\bk}g^*_{\bk})^2
\end{align}
 we have 
\begin{align}
\omega_{\bk}^2(\zeta)=& C^{-1}(\zeta)(\bk^2+C(\zeta)\mKG^2)
\end{align}

\subsection{Schr\"{o}dinger 2nd Quantized Hamiltonian}
In Schr\"{o}dinger picture, canonical fields are constant and agree with Heisenberg pictures fields at the chosen time $\zeta=\ell$. The Hamiltonian is then eq.(\ref{SchHa})
with
\begin{align}
\phi^{\text{S}}(\bx)=&\sum_{\bk}(a_{\bk}\Phi_{\bk}(\ell,\bx)+a^\dag_{\bk}\Phi^*_{\bk}(\ell,\bx))\\
\pi^{\text{S}}(\bx)=& \sum_{\bp}(a_{\bp}\nabla_\zeta\Phi_{\bp}(\ell,\bx)+a^\dag_{\bp}\nabla_\zeta\Phi^*_{\bp}(\ell,\bx))
\end{align}
Thus, we have 
\begin{align}
H^{\text{S}}
&=\frac{1}{2}C^{3/2}(\zeta)(2\pi)^3\sum_{\bk} [C^{-3}(\zeta)(\dot{g}_{-\bk}(\ell)\dot{g}_{\bk}(\ell)a_{-\bk}a_{\bk}\nonumber\\
&+a_{\bk}a^\dag_{\bk}\dot{g}_{\bk}(\ell)\dot{g}^*_{\bk}(\ell)+a^\dag_{\bk}a_{\bk}\dot{g}^*_{\bk}(\ell)\dot{g}_{\bk}(\ell)+a^\dag_{-\bk}a^\dag_{\bk}\dot{g}^*_{-\bk}(\ell)\dot{g}^*_{\bk}(\ell))\nonumber\\
&+(C^{-1}(\zeta)\bk^2+\mKG^2)(a_{-\bk}a_{\bk}g_{-\bk}(\ell)g_{\bk}(\ell)+a_{\bk}a^\dag_{\bk}g_{\bk}(\ell)g^*_{\bk}(\ell)\nonumber\\
&+a_{\bk}^\dag a_{\bk}g^*_{\bk}(\ell)g_{\bk}(\ell)+a^\dag_{-\bk}a^\dag_{\bk}g^*_{-\bk}(\ell)g^*_{\bk}(\ell))]
\end{align}   
As in Heisenberg picture, we write
  \begin{align}
H^{\text{S}}(\zeta)&=\frac{1}{2}\sum_{\bk}\vep^{\text{S}}_{\bk}(\zeta)(a^\dag_{\bk} a_{\bk}+a_{\bk}a^\dag_{\bk})\nonumber\\
&+\frac{1}{2}\sum_{\bk}\left[\Delta^{\text{S}}_{\bk}(\zeta)a_{\bk}a_{-\bk}+\Delta^{\text{S}*}_{\bk}(\zeta)a^\dag_{-\bk}a^\dag_{\bk}\right]
\end{align}
where
\begin{align}
\vep^{\text{S}}_{\bk}
=&(2\pi)^3C^{1/2}(\zeta)\left[\frac{|\dot{g}_{\bk}(\ell)|^2}{C^2(\zeta)}+\left(\bk^2+C(\zeta)\mKG^2\right)|g_{\bk}(\ell)|^2\right]\\
\Delta^{\text{S}}_{\bk}
=&(2\pi)^3C^{1/2}(\zeta)\left[C^{-2}(\zeta)\dot{g}_{\bk}^2(\ell)+\left(\bk^2+C(\zeta)\mKG^2\right)g_{\bk}^2(\ell)\right]
\end{align}   
By the same token, one can calculate
\begin{align}
\omega^{\text{S}2}_{\bk}(\zeta)=&-(2\pi)^6C^{-1}(\zeta)\left(\bk^2+C(\zeta)\mKG^2\right)\nonumber\\
&\times(\dot{g}_{\bk}(\ell)^* g_{\bk}(\ell)-\dot{g}_{\bk}(\ell)g^*_{\bk}(\ell))^2
\end{align}
At $\zeta=\ell,  \dot{g}_{\bk}^*(\ell) g_{\bk}(\ell)-\dot{g}_{\bk}(\ell)g^*_{\bk}(\ell)=(2\pi)^{-3}(-i) $. Hence $\omega^{\text{S}2}_{\bk}(\zeta)=\omega^2_{\bk}
$. The Bogliubov transformation is
\begin{align}
b^{\text{S}}_{\bk}(\zeta)=&u^{\text{S}}_{\bk}(\zeta) a_{\bk}+v^{\text{S}}_{\bk}(\zeta) a_{-\bk}^{\dag}\\
b^{\text{S}\dag}_{\bk}(\zeta)=&u^{\text{S}*}_{\bk}(\zeta) a_{\bk}^{\dag}+v^{\text{S}*}_{\bk}(\zeta) a_{-\bk}
\end{align}
with
\begin{align}
u^{\text{S}}_{\bk}=&\sqrt{\frac{\vep^{\text{S}}_{\bk}+\omega_{\bk}(\zeta)}{2\omega_{\bk}(\zeta)}} ,  \quad v^{\text{S}}_{\bk}=\frac{\Delta^{\text{S}*}_{\bk}}{\vep^{\text{S}}_{\bk}+\omega_{\bk}(\zeta)}u^{\text{S}}_{\bk}
\end{align}
The Schr\"{o}dinger picture Hamiltonian is 
\beq
H^{\text{S}}(\zeta)=\frac{1}{2}\sum_{\bk}\omega_{\bk}(\zeta)(b^{\text{S}\dag}_{\bk}(\zeta)b^{\text{S}}_{\bk}(\zeta)+b^{\text{S}}_{\bk}(\zeta)b^{\text{S}\dag}_{\bk}(\zeta))\not=H(\zeta)
\eeq
The relation between  quasi-creation/annihilation operators $b_{\bk}(\zeta), b^{\text{S}}_{\bk}(\zeta)$ is
\begin{align}
b^{\text{S}}_{\bk}(\zeta)=&\left[u^{\text{S}}(\zeta)u^*_{\bk}(\zeta)-v^{\text{S}}_{\bk}(\zeta)v^*_{\bk}(\zeta)\right]b_{\bk}(\zeta)\nonumber\\
&
+\left[v^{\text{S}}(\zeta)u_{\bk}(\zeta)-u^{\text{S}}_{\bk}(\zeta)v_{\bk}(\zeta)\right]b^\dag_{-\bk}(\zeta)\\
b^{\text{S}\dag}_{\bk}(\zeta)=&\left[u^{\text{S}*}(\zeta)u_{\bk}(\zeta)-v^{\text{S}*}_{\bk}(\zeta)v_{\bk}(\zeta)\right]b^\dag_{\bk}(\zeta)\nonumber\\
&
+\left[v^{\text{S}*}(\zeta)u^*_{\bk}(\zeta)-u^{\text{S}*}_{\bk}(\zeta)v^*_{\bk}(\zeta)\right]b_{-\bk}(\zeta)
\end{align}
and the inverse is
\begin{align}
b_{\bk}(\zeta)=&\left[u(\zeta)u^{\text{S}*}_{\bk}(\zeta)-v_{\bk}(\zeta)v^{\text{S}*}_{\bk}(\zeta)\right]b^{\text{S}}_{\bk}(\zeta)\nonumber\\
&+\left[v(\zeta)u^{\text{S}}_{\bk}(\zeta)-u_{\bk}(\zeta)v^{\text{S}}_{\bk}(\zeta)\right]b^{\text{S}\dag}_{-\bk}(\zeta)\\
b^\dag_{\bk}(\zeta)=&\left[u^*(\zeta)u^{\text{S}}_{\bk}(\zeta)-v^*_{\bk}(\zeta)v^{\text{S}}_{\bk}(\zeta)\right]b^{\text{S}\dag}_{\bk}(\zeta)\nonumber\\
&+\left[v^*(\zeta)u^{\text{S}^*}_{\bk}(\zeta)-u^*_{\bk}(\zeta)v^{\text{S}*}_{\bk}(\zeta)\right]b^{\text{S}}_{-\bk}(\zeta)
\end{align}
These relations are important to perturbative calculations, as will be discussed later on. The measured energy-momentum is\cite{Feng2001}
\beq
p_a(\zeta)=(\omega_{\bk}, C^{-1/2}(\zeta)\bk)
\eeq
So we have on-shell relation of measured 4-momentum $p_a(\zeta)p^a(\zeta)=\mKG^2$, as in Minkowski spacetime.  We have accordingly the red-shift relation
\beq
\frac{\omega_{\bk}(\zeta_1)}{\omega_{\bk}(\zeta_2)}=\sqrt{\frac{C^{-1}(\zeta_1)(\bk^2+C(\zeta_1)\mKG^2)}{C^{-1}(\zeta_2)(\bk^2+C(\zeta_2)\mKG^2)}}
\eeq
This relation seems new to our best knowledge.  Since many existing discussions suffer lacking both unique and invariant definition of observable quantum particle states, concepts such as energy are not well defined.   The impact of curved spacetime on measured energy of particles was discussed via phase analysis \cite{Hochberg1991} and the behavior of $\omega_{\bk}(\zeta)$ here agrees exactly with the  the geometric-optics limit  \cite{Feng2001}.
This redshift relation agrees with the conventional relation $\omega_B/\omega_A=\sqrt{g_{00}(x_A)/g_{00}(x_B)}$ in the zero-mass limit. 

Similar to the vacuum state of Fock space of operators $a_{\bk}, a^\dag_{\bk}$
\beq
|0\rj=\prod_{\bk}|0_{\bk}\rj, a_{\bk}|0_{\bk}\rj=0
\eeq
we definie $|\bO;\zeta\rj^{\text{S}}$ s.t.
\beq
 b^{\text{S}}_{\bk}(\zeta)|\bO;\zeta\rj^{\text{S}}=0,\,\,\,\,\, \forall \,\,\,\,\ \bk,
\eeq
i.e.
\beq
|\bO;\zeta\rj^{\text{S}}=\prod_{\bk} |\bO_{\bk};\zeta\rj^{\text{S}} _{\bk}
\eeq
where
\beq
b^{\text{S}}_{\bk} |\bO_{\bk};\zeta\rj_{\bk}^{\text{S}}=0.
\eeq
 Since
\beq
\lj 0|H(\zeta)|0\rj=\sum_{\bk}\vep_{\bk}>\,^{\text{S}}\lj\bO;\zeta| H(\zeta)|\bO;\zeta\rj^{\text{S}}=\sum_{\bk}\omega_{\bk}(\zeta)
\eeq
,$|0\rj$ is not the ground-state.  Here we have a system of time-dependent oscillators, as the cases discussed in \cite{Chernikov1967}-\cite{Struckmeier2001} . The vacuum states are time-dependent.  The vacuum state at one time will evolve into a non-vacuum state at a later time. Thus particles can be generated.

In the rest part of this section and the next, states and generation/annihilation operators $b^\dag_{\bk}(\zeta), b_{\bk}(\zeta) $ can be in either Schr\"{o}dinger picture or Heisenberg pictures since the math structures are essentially the same.
 Denoting 
\begin{align}
|\{n_{\bk}\};\zeta\rj
=\prod_{i} \Big\{\frac{1}{\sqrt{n_{\bk_i}!}}\left[b_{\bk_i}^\dag(\zeta)\right]^{ n_{\bk_i}}|\bO_{k_i};\zeta\rj_{\bk_i}\Big\}
\end{align}
 and $|1_{\bk};\zeta\rj=|\bk;\zeta\rj$.  
Using results in \cite{Miransky1993}, we have
\beq
|\bO;\zeta\rj=\hat{A}(\zeta)|0\rj=\otimes_{\bk}|\bO_{\bk};\zeta\rj_{\bk}
\eeq
where
\beq
\hat{A}(\zeta)=\mathpzc{A}(\zeta)e^{-\frac{1}{2}\sum_{\bk}\frac{v_{\bk}(\zeta)}{u_{\bk}(\zeta)}a^\dag_{-\bk}a^\dag_{\bk}}
\eeq
with
\beq
\mathpzc{A}(\zeta)=
\prod_{\bk}\,' \mathpzc{A}_{\bk}(\zeta)
\eeq
where $\prod_{\bk}\,'$ means that only one of the two factors corresponding to $\bk,-\bk$ appears in the product and
$
\mathpzc{A}_{\bk}(\zeta)=1/u_{\bk}(\zeta). 
$
Denoting
\beq
\gamma_{\bp}(\zeta)=\frac{v_{\bp}(\zeta)}{u_{\bp}(\zeta)}=\frac{\theta^*_{\bk}(\zeta)}{|\theta^*_{\bk}(\zeta)|}\text{th}|\theta_{\bk}(\zeta)|
\eeq
then
\beq
\hat{A}(\zeta)=\mathpzc{A}(\zeta)e^{-\frac{1}{2}\sum_{\bk}\gamma_{\bk}(\zeta)a^\dag_{-\bk}a^\dag_{\bk}}=\prod_{\bk}\,' \hat{A}_{\bk}(\zeta)
\eeq
with
\beq
\hat{A}_{\bk}(\zeta)=\mathpzc{A}_{\bk}(\zeta)e^{-\gamma_{\bk}(\zeta)a^\dag_{-\bk}a^\dag_{\bk}}=\hat{A}_{-\bk}(\zeta)
\eeq
For a particlular $\bk$
\beq
|\bO_{\bk};\zeta\rj_{\bk}=\mathpzc{A}_{\bk}(\zeta)e^{-\gamma_{\bk}(\zeta)a^\dag_{-\bk}a^\dag_{\bk}}|0\rj_{\bk} \otimes|0\rj _{-\bk}=|\bO_{-\bk};\zeta\rj_{-\bk}
\eeq
We can calculate  $\lj\bO;\zeta_1| \bO;\zeta_2\rj$.  First,  we have
\begin{align}
_{\bk}\lj \bO_{\bk};\zeta_1|\bO_{\bk};\zeta_2\rj_{\bk}
=&\frac{1}{u_{\bk}(\zeta_1)u_{\bk}(\zeta_2)-v^*_{\bk}(\zeta_1)v_{\bk}(\zeta_2)}
\end{align}
Then
\begin{align}
\lj \bO;\zeta_1|\bO;\zeta_2\rj
=\prod_{\bk}\frac{1}{\sqrt{u_{\bk}(\zeta_1)u_{\bk}(\zeta_2)-v^*_{\bk}(\zeta_1)v_{\bk}(\zeta_2)}}  \label{O|O}
\end{align}
Using the following expectations
\begin{align}
&\lj\bO;\zeta''|a_{\bk}a_{\bp}|\bO;\zeta'\rj \nonumber\\
=&\left\{\begin{array}{cl}
   0, & \bp\not=\pm\bk , \bp=\bk\\
\lj\bO;\zeta''|\bO;\zeta'\rj \frac{-1}{1-\gamma^*_{\bk}(\zeta'')\gamma_{\bk}(\zeta')}, & \bp=-\bk\\
       \end{array}\right.
\end{align}
\begin{align}
&\lj\bO;\zeta''|a_{\bk}a^\dag_{\bp}|\bO;\zeta'\rj \nonumber\\
=&\left\{\begin{array}{cl}
   0, & \bp\not=\pm\bk , \bp=-\bk\\
\lj\bO;\zeta''|\bO;\zeta'\rj \frac{1}{1-\gamma^*_{\bk}(\zeta'')\gamma_{\bk}(\zeta')}, & \bp=\bk\\
       \end{array}\right.
\end{align}
one can calculate 
\begin{align}
&\lj\bO;\zeta''|b_{\bk}(\zeta'')b^\dag_{\bk}(\zeta')|\bO;\zeta'\rj \nonumber\\
=&\lj\bO;\zeta''|\bO;\zeta'\rj\times \,
_{\bk}\lj\bO_{\bk};\zeta''|\bO_{\bk};\zeta'\rj_{\bk}
\end{align}
Hence
\begin{align}
_{\bk}\lj\bO_{\bk};\zeta''|b_{\bk}(\zeta'')b^\dag_{\bk}(\zeta')|\bO_{\bk};\zeta'\rj_{\bk}
=\left[_{\bk}\lj\bO_{\bk};\zeta''|\bO_{\bk};\zeta'\rj_{\bk}\right]^2
\end{align}
Similarly
\begin{align}
&_{\bk}\lj\bO_{\bk};\zeta''|b_{\bk}(\zeta'')b_{-\bk}(\zeta'')|\bO_{\bk};\zeta'\rj _{\bk}\nonumber\\
 =&\frac{_{\bk}\lj\bO_{\bk};\zeta''|\bO_{\bk};\zeta'\rj_{\bk}}{1-\gamma^*_{\bk}(\zeta'')\gamma_{\bk}(\zeta')}
\left[
\gamma_{\bk}(\zeta'')-\gamma_{\bk}(\zeta')
\right]
\label{bb0}
\end{align}
\section{Generation Functional of Vacuum Expectations}
We define generation functional
\begin{align}
&\mathpzc{Z}[\lambda,\lambda^*;\zeta,\zeta']\nonumber\\
:=&\prod_{\bk}\,'_{\bk}\lj \bO_{\bk};\zeta|
e^{\int d\eta\lambda_{\bk}(\eta)b_{\bk}(\eta)}e^{\int d\eta\lambda_{-\bk}(\eta)b_{-\bk}(\eta)}\nonumber\\
&\times e^{\int d\eta'(\lambda^*_{\bk}(\eta')b_{\bk}^\dag(\eta')}e^{\int d\eta'\lambda^*_{-\bk}(\eta')b_{-\bk}^\dag(\eta')}|\bO_{\bk};\zeta'\rj_{\bk}
\end{align}
The $\lambda_{\bk}(\eta), \lambda^*_{\bk}(\eta)$ are considered as external fields and are independent of each other, instead of being mutually complex conjugate . For a particular $\bk$, we define
\begin{align}
&\mathpzc{Z}_{\bk}[\lambda,\lambda^*;\zeta,\zeta']\nonumber\\
:=&\lj \bO_{\bk};\zeta|
e^{\int d\eta\lambda_{\bk}(\eta)b_{\bk}(\eta)}e^{\int d\eta\lambda_{-\bk}(\eta)b_{-\bk}(\eta)}\nonumber\\
&\times 
e^{\int d\eta'(\lambda^*_{\bk}(\eta')b_{\bk}^\dag(\eta')}e^{\int d\eta'\lambda^*_{-\bk}(\eta')b_{-\bk}^\dag(\eta')}|\bO_{\bk};\zeta'\rj_{\bk}\nonumber\\
=&_{\bk}\lj \bO_{\bk};\zeta| 
e^{\int d\eta\lambda_{\bk}(\eta)\left[u_{\bk}(\eta) a_{\bk}+v_{\bk}(\eta) a_{-\bk}^{\dag}\right]}\nonumber\\
&\times e^{\int d\eta\lambda_{-\bk}(\eta)\left[u_{-\bk}(\eta) a_{-\bk}+v_{-\bk}(\eta) a_{\bk}^{\dag}\right]}\nonumber\\
&\times
e^{\int d\eta'\lambda^*_{\bk}(\eta')\left[u_{\bk}^{*}(\eta') a_{\bk}^{\dag}+v_{\bk}^{*}(\eta') a_{-\bk}\right]}\nonumber\\
&\times
e^{\int d\eta'\lambda^*_{-\bk}(\eta')\left[u_{-\bk}^{*}(\eta') a_{-\bk}^{\dag}+v_{-\bk}^{*}(\eta') a_{\bk}\right]}|\bO_{\bk};\zeta'\rj_{\bk}
\end{align}
Denoting
\beq
\alpha_{\bk}[\lambda]=\int d\eta \lambda_{\bk}(\eta)u_{\bk}(\eta),\,\,\,\,\,\,\,  \beta_{\bk}[\lambda]=\int d\eta \lambda_{\bk}(\eta)v_{\bk}(\eta),
\eeq
\begin{align}
\mathpzc{Z}_{\bk}[\lambda,\lambda^*;\zeta,\zeta']
=&_{\bk}\lj \bO_{\bk};\zeta| 
e^{\alpha_{\bk}a_{\bk}}\cdot e^{\beta_{\bk}a^\dag_{-\bk}}\cdot e^{\alpha_{-\bk}a_{-\bk}}\cdot e^{\beta_{-\bk}a^\dag_{\bk}}\nonumber\\
\times &
 e^{\alpha^*_{\bk}a^\dag_{\bk}}\cdot e^{\beta^*_{\bk}a_{-\bk}}\cdot e^{\alpha^*_{-\bk}a^\dag_{-\bk}}\cdot e^{\beta^*_{-\bk}a_{\bk}}
|\bO_{\bk};\zeta'\rj_{\bk}
\end{align}
Using coherent states $|z_{\bk},z_{-\bk}\rj=e^{ z_{\bk}a^\dag_{\bk}+z_{-\bk}a^\dag_{-\bk}}|0\rj$, one can calculate
\begin{align}
&  \mathpzc{Z}_{\bk}[\lambda,\lambda^*;\zeta,\zeta']\nonumber\\
=&\frac{\mathpzc{A}_{\bk}(\zeta')\mathpzc{A}_{\bk}^*(\zeta)}{1-\gamma^*_{\bk}(\zeta)\gamma_{\bk}(\zeta')}\nonumber\\
&\times 
\exp\Big[-\beta^*_{-\bk}(\alpha^*_{\bk}+\beta_{-\bk})-\beta_{\bk}(\alpha_{-\bk}+\beta^*_{\bk})\nonumber\\
&+\frac{1}{1-\gamma^*_{\bk}(\zeta)\gamma_{\bk}(\zeta')}(A^*_{\bk}A_{\bk}+A^*_{-\bk}A_{-\bk}\nonumber\\
&-A^*_{-\bk}\gamma_{\bk}(\zeta')A^*_{\bk}-A_{\bk}\gamma^*_{\bk}(\zeta)A_{-\bk})\Big]
\end{align}
, with which the following amplitudes can be evaluated
\begin{align}
_{\bk}\lj 1_{\bk};\zeta|1_{\bk};\zeta'\rj_{\bk}=&\frac{\delta}{\delta \lambda_{\bk}(\zeta)}\frac{\delta}{\delta \lambda^*_{\bk}(\zeta')}
\mathpzc{Z}_{\bk}[\lambda,\lambda^*;\zeta,\zeta']_{|\lambda=\lambda^*=0}\nonumber\\
=&\frac{1}{\left[u_{\bk}(\zeta)u_{\bk}(\zeta')(1-\gamma^*_{\bk}(\zeta)\gamma_{\bk}(\zeta')\right]^2}
\end{align}
and
\begin{align}
&_{\bk}\lj 1_{\bk}1_{-\bk}; \zeta|\bO;\zeta'\rj_{\bk}\nonumber\\
=&\frac{\delta}{\delta \lambda_{\bk}(\zeta)}\frac{\delta}{\delta \lambda_{-\bk}(\zeta)}
\mathpzc{Z}_{\bk}[\lambda,\lambda^*;\zeta,\zeta']_{|\lambda=\lambda^*=0}\nonumber\\
=&\frac{\mathpzc{A}_{\bk}(\zeta')\mathpzc{A}_{\bk}^*(\zeta)}{\left[1-\gamma^*_{\bk}(\zeta)\gamma_{\bk}(\zeta'\right]^2}
(\gamma(\zeta)-\gamma(\zeta'))
\end{align}
\section{Transition of States of Free KG Field}
 As in conventional quantum field theories\cite{Itzykson1980}, suppose that at time $\zeta_1$, the system is in a vacuum state, the state at $\zeta_2$ has some computable probability to contain multiple particles. In Schr\"{o}dinger picture, the state
\begin{align}
|\Psi_0(\zeta_2;\zeta_1)\rj^{\text{S}}=\hat{T}e^{-i\int^{\zeta_2}_{\zeta_1}H^{\text{S}}[\phi(\bx),\pi(\bx);\eta]e^0_\zeta(\eta)d\eta}|\bO;\zeta_1\rj ^{\text{S}}
\end{align}
is a formal solution to
\beq
i\hat{\nabla}_0|\Psi_0(\zeta;\zeta_1)\rj^{\text{S}}=H^{\text{S}}[\phi(\bx),\pi(\bx);\zeta] |\Psi_0(\zeta;\zeta_1)\rj ^{\text{S}}
\eeq
with initial condition $|\Psi_0(\zeta_1;\zeta_1)\rj ^{\text{S}}=|\bO;\zeta_1\rj ^{\text{S}}$.
The transition amplitude from state $|\{n_{\bk}\};\zeta_1\rj^{\text{S}}$ to state  $|\{m_{\bk}\};\zeta_2\rj ^{\text{S}}$ is given by
\begin{align}
&\mathscr{T}(|\{n_{\bk}\};\zeta_1\rj ^{\text{S}}\rightarrow |\{m_{\bk}\};\zeta_2\rj ^{\text{S}})\nonumber\\
=&\,^{\text{S}}\lj\{m_{\bk}\};\zeta_2| \hat{T}e^{-i\int^{\zeta_2}_{\zeta_1}H^{\text{S}}(\eta)e^0_\zeta(\eta)d\eta}|\{n_{\bk}\};\zeta_1\rj ^{\text{S}}
\end{align}
For a particular $\bk$, 
\begin{align}
&\mathscr{T}(|m_{\bk},m_{-\bk};\zeta_1\rj_{\bk}^{\text{S}}\rightarrow |n_{\bk},n_{-\bk};\zeta_2\rj_{\bk}^{\text{S}})\nonumber\\
&=\,^{\text{S}}_{\bk}\lj n_{\bk},n_{-\bk};\zeta_2|\hat{T}e^{-i\int^{\zeta_2}_{\zeta_1}\omega_{\bk}(\eta)\left[b^{\text{S}\dag}_{\bk}(\eta)b^{\text{S}}_{\bk}(\eta)+
b^{\text{S}\dag}_{-\bk}(\eta)b^{\text{S}}_{-\bk}(\eta)+1\right]e^0_\zeta(\eta) d\eta}\nonumber\\
&\times |m_{\bk},m_{-\bk};\zeta_1\rj_{\bk} ^{\text{S}}
\end{align}
Slicing the time as 
$\eta_0=\zeta_1, \eta_{j}=\eta_{j-1}-\varDelta,\varDelta=\frac{\zeta_1-\zeta_2}{N}, \eta_N=\zeta_2$. 
\begin{align}
&\mathscr{T}(|m_{\bk},m_{-\bk};\zeta_1\rj_{\bk}^{\text{S}}\rightarrow |n_{\bk},n_{-\bk};\zeta_2\rj_{\bk}^{\text{S}})\nonumber\\
=&\lim_{N\rightarrow\infty}\, ^{\text{S}}_{\bk}\lj n_{\bk},n_{-\bk};\zeta_2|\nonumber\\
&\times
e^{-i\omega_{\bk}(\eta_{N-1})\left[b^{\text{S}\dag}_{\bk}(\eta_{N-1})b^{\text{S}}_{\bk}(\eta_{N-1})+b^{\text{S}\dag}_{-\bk}(\eta_{N_1})b^{\text{S}}_{-\bk}(\eta_{N-1})+1\right]e^0_\zeta(\eta_{N-1}) \varDelta}\nonumber\\
&\times
e^{-i\omega_{\bk}(\eta_{N-2})\left[b^{\text{S}\dag}_{\bk}(\eta_{N-2})b^{\text{S}}_{\bk}(\eta_{N-2})+b^{\text{S}\dag}_{-\bk}(\eta_{N_2})b^{\text{S}}_{-\bk}(\eta_{N-2})+1\right]e^0_\zeta(\eta_{N-2}) \varDelta}
\nonumber\\
&\times \cdots
e^{i\omega_{\bk}(\eta_{1})\left[b^{\text{S}\dag}_{\bk}(\eta_{1})b^{\text{S}}_{\bk}(\eta_{1})+b^{\text{S}\dag}_{-\bk}(\eta_{1})b^{\text{S}}_{-\bk}(\eta_{1})+1\right]e^0_\zeta(\eta_{1})\varDelta }\nonumber\\
&\times 
e^{-i\omega_{\bk}(\eta_{0})\left[b^{\text{S}\dag}_{\bk}(\eta_{0})b^{\text{S}}_{\bk}(\eta_{0})+b^{\text{S}\dag}_{-\bk}(\eta_{0})b^{\text{S}}_{-\bk}(\eta_{0})+1\right]e^0_\zeta(\eta_{0})\varDelta }
|m_{\bk},m_{-\bk};\zeta_1\rj_{\bk}^{\text{S}}
\end{align}
Using coherent states $j=0,1,\cdots, N$
\beq
|z_{j\bk}, z_{j-\bk}\rj=e^{z_{j\bk}b^{\text{S}\dag}_{\bk}(\eta_j)+z_{j-\bk}b^{\text{S}\dag}_{-\bk}(\eta_j)}|\bO_{\bk};\eta_j\rj_{\bk}
\eeq
\begin{align}
&\mathscr{T}(|m_{\bk},m_{-\bk};\zeta_1\rj_{\bk}^{\text{S}}\rightarrow |n_{\bk},n_{-\bk};\zeta_2\rj_{\bk}^{\text{S}})\nonumber\\
&=\lim_{N\rightarrow\infty}e^{-i\int^{\zeta_2}_{\zeta_1}\omega_{\bk}(\eta)e^0_{\zeta}(\eta)d\eta}
\int \left[\prod_{j=0}^N\frac{dz_{j\bk}dz^*_{j\bk}}{2\pi i} \right]\nonumber\\
&\times  \frac{z^{n_{\bk}}_{N\bk}}{\sqrt{n_{\bk}!}}
  \frac{z^{*m_{\bk}}_{0\bk}}{\sqrt{m_{\bk}!}} 
\exp\left[-z^*_{0\bk}z_{0\bk}\right]\nonumber\\
&\times
\prod_{j=1}^{N}
\exp\left[-z^*_{j\bk}z_{j\bk}+z^*_{j\bk}z_{j-1,\bk}-i\omega_{\bk}(\eta_{j-1})z^*_{j\bk}z_{j-1\bk}e^0_\zeta(\eta_{j-1}) \varDelta\right]\nonumber\\
&\times
\int \left[\prod_{j=0}^N \frac{dz_{j-\bk}dz^*_{j-\bk}}{2\pi i}\right]\,
 \frac{z^{n_{-\bk}}_{N-\bk}}{\sqrt{n_{-\bk}!}}
  \frac{z^{*m_{-\bk}}_{0-\bk}}{\sqrt{m_{-\bk}!}} 
\exp\left[-z^*_{0-\bk}z_{0-\bk}\right]\nonumber\\
&\times
\prod_{j=1}^{N}
\exp\Big[-z^*_{j-\bk}z_{j-\bk}+z^*_{j-\bk}z_{j-1,-\bk}\nonumber\\
&-i\omega_{\bk}(\eta_{j-1})z^*_{j-\bk}z_{j-1,-\bk}e^0_\zeta(\eta_{j-1}) \varDelta\Big]
\end{align}
which can be written as a path-integral by shorthand. It is easy to see that starting from a vacuum state at $\zeta_1$, the state will evolve into a mixed states at later time $\zeta_2$, which is not unusual for systems in external fields in  Minkowski quantum field theories\cite{Itzykson1980} and time-dependent harmonic oscillators\cite{Chernikov1967}-\cite{Struckmeier2001}.
\section{Perturbation}
The full Lagrangian of $\lambda^4$-theory is
\beq
\mathscr{L}=\frac{1}{2}(\hat{\nabla}^a\phi\hat{\nabla}_a\phi-\mKG^2\phi^2-\frac{\lambda}{4!}\phi^4)
\eeq
and the full Hamiltonian is
\begin{align}
\mathscr{H}=&\hat{\nabla}_0 \phi \cdot\Pi-\mathscr{L}=\mathscr{H}_0+\mathscr{H}_{\text{I}}\\
H[\phi,\Pi;\zeta]=&H_0[\phi,\Pi;\zeta]+H_{\text{I}}[\phi,\Pi;\zeta]
\end{align}
where
\begin{align}
\mathscr{H}_0=&\frac{1}{2}(\Pi^2-\hat{\nabla}^{a'}\phi\hat{\nabla}_{a'}\phi+\mKG^2\phi^2)\\
\mathscr{H}_{\text{I}}=&\frac{1}{2}\frac{\lambda}{4!}\phi^4
\end{align}
 The Schr\"{o}dinger state follows
 \beq
 i\hat{\nabla}_0|\Psi(\zeta)\rj^{\text{S}}=H[\phi^{\text{S}}(\bx),\pi^{\text{S}}(\bx);\zeta] |\Psi(\zeta)\rj^{\text{S}}
  \eeq
Defining Dirac picture state  
 \beq
|\Psi(\zeta)\rj^{\text{D}}:=\hat{T}^{-1}e^{i\int_\ell^{\zeta}H_0^{\text{S}}(\eta)e^0_\zeta(\eta)d\eta} |\Psi(\zeta)\rj^{\text{S}}
  \eeq  
  hence the two pictures coincide at time $\zeta=\ell,
|\Psi(\ell)\rj^{\text{S}}=|\Psi(\ell)\rj^{\text{D}}
$
  we thus have equation of motion 
 \begin{align}
 i\hat{\nabla}_0| \Psi(\zeta)\rj^{\text{D}}
=&H^{\text{D}}_{\text{I}}(\zeta)  |\Psi(\zeta)\rj^{\text{D}}
  \end{align} 
  where
\begin{align}
 H^{\text{D}}_{\text{I}}(\zeta)
=&U^{\text{S}-1}_0(\zeta,\ell) H_{\text{I}}^{\text{S}}(\zeta)U^{\text{S}}_0(\zeta,\ell)
\end{align}
with
\beq
U^{\text{S}}_0(\zeta'',\zeta')=\hat{T} e^{-i\int^{\zeta''}_{\zeta'} H^{\text{S}}_0(\eta)e^0_\zeta(\eta)d\eta}
\eeq
The Dirac picture field and operators are defined
\begin{align}
\phi^{\text{D}}(\zeta,\bx):=&U^{\text{S}-1}_0(\zeta,\ell)\phi^{\text{S}}(\bx)U^{\text{S}}_0(\zeta,\ell)\\
H_0^{\text{D}}(\zeta):=&U^{\text{S}-1}_0(\zeta,\ell) H_0^{\text{S}}(\zeta)U^{\text{S}}_0(\zeta,\ell) \label{HD}
\end{align}
, from which it follows that
\begin{align}
ie_0^{\zeta}(\zeta)\frac{\p}{\p\zeta} \phi^{\text{D}}(\zeta,\bx)=&[\phi^{\text{D}}(\zeta,\bx),H^{\text{D}}_0]\\
ie_0^{\zeta}(\zeta)\frac{\p}{\p\zeta} \pi^{\text{D}}(\zeta,\bx)=&[\pi^{\text{D}}(\zeta,\bx),H^{\text{D}}_0]
\end{align}
Hence $\phi^{\text{D}}(\zeta,\bx),\pi^{\text{D}}(\zeta,\bx)$ follow equation of motion of a non-interacting field, of which the time-dependence was discussed previously.
Defining
\begin{align}
U^{\text{D}}_0(\zeta'',\zeta')=&\hat{T} e^{-i\int^{\zeta''}_{\zeta'} H^{\text{D}}_0(\eta)e^0_\zeta(\eta)d\eta}
\end{align}
we have
\begin{align}
U^{\text{D}\dag}_0(\zeta'',\zeta')=&\hat{T}^{-1} e^{i\int^{\zeta''}_{\zeta'} H^{\text{D}}_0(\eta)e^0_\zeta(\eta)d\eta}=U^{\text{D}-1}_0(\zeta'',\zeta')
\end{align}
Further, by definition eq.(\ref{HD}) of $H_0^{\text{D}}(\zeta)$, we have
\beq
H_0^{\text{D}}(\zeta)e^0_\zeta(\zeta)=U^{\text{S}-1}_0(\zeta,\ell) i\p_\zeta U^{\text{S}}_0(\zeta,\ell)
\eeq
Therefore
\beq
i\p_\zeta U^{\text{S}-1}_0(\zeta,\ell)=-H_0^{\text{D}}(\zeta)e^0_\zeta(\zeta) U^{\text{S}-1}_0(\zeta,\ell)
\eeq
Hence we have the following relations
\begin{align}
U^{\text{S}-1}_0(\zeta,\ell)= &\hat{T}e^{i\int^{\zeta}_{\ell} H^{\text{D}}_0(\eta)e^0_\zeta(\eta)d\eta}\\
U^{\text{S}}_0(\zeta,\ell)= &\hat{T}^{-1}e^{-i\int^{\zeta}_{\ell} H^{\text{D}}_0(\eta)e^0_\zeta(\eta)d\eta}\\
\hat{T}e^{-i\int_\ell^{\zeta}H_0^{\text{S}}(\eta)e^0_\zeta(\eta)d\eta}=&\hat{T}^{-1}e^{-i\int_\ell^{\zeta}H_0^{\text{D}}(\eta)e^0_\zeta(\eta)d\eta}
\end{align}
we have inverse transformation
\beq
H_0^{\text{S}}(\zeta)=\hat{T}^{-1}e^{-i\int_\ell^{\zeta}H_0^{\text{D}}(\eta)e^0_\zeta(\eta)d\eta} H_0^{\text{D}}(\zeta)\hat{T}e^{i\int_\ell^{\zeta}H_0^{\text{D}}(\eta)e^0_\zeta(\eta)d\eta}
\eeq
and
\beq
| \Psi(\zeta)\rj^{\text{S}}
=\hat{T}^{-1}e^{-i\int_\ell^{\zeta}H_0^{\text{D}}(\eta)e^0_\zeta(\eta)d\eta}| \Psi(\zeta)\rj^{\text{D}}
  \eeq  

Suppose at the initial time $\zeta=\ell$, the system is in the eigen-state $|A;\ell\rj_0^{\text{S}}$ of $H_0^{\text{S}}(\ell)$, then interaction is turned on adiabatically.
At time $\zeta=0$, the interaction is turned off and the state evolves  into a state which can be expanded in terms of eigen-states $\{|B;0\rj_0^{\text{S}}\} $ of $H_0^{\text{S}}(0)$. The prob of the transition is the square of the amplitude $^{\text{S}}_0\lj B;0| U^{\text{S}}(0,\ell)|A,\ell\rj_0^{\text{S}}$
 \cite[p.323]{Merzbacher1998}, where the Schr\"{o}dinger picture evolution operator is
\beq
U^{\text{S}}(\zeta_2,\zeta_1)=\hat{T}e^{-i\int_{\zeta_1}^{\zeta_2}H^{\text{S}}(\eta)e^0_{\zeta}(\eta)d\eta}
\eeq
For a free field, we have
\beq
^{\text{S}}_0\lj B;0| U^{\text{S}}(0;\ell)|A,\ell\rj_0^{\text{S}}=\,^{\text{S}}_0\lj B;0| \hat{T}e^{-i\int_\ell^{0}H_0^{\text{S}}(\eta)e^0_{0}(\eta)d\eta}|A,\ell\rj_0^{\text{S}}
\eeq
, which was discussed previously. In the interacting case, in terms of Dirac picture
\beq
| \Psi(\zeta)\rj^{\text{D}}
=\hat{T}e^{-i\int_\ell^{\zeta}H^{\text{D}}_{\text{I}}(\eta)e^0_{\zeta}(\eta)d\eta}  |\Psi(\ell)\rj^{\text{D}}
  \eeq  
\begin{align}
&| \Psi(\zeta)\rj^{\text{S}}\nonumber\\
=&\hat{T}^{-1}e^{-i\int_\ell^{\zeta}H_0^{\text{D}}(\eta)e^0_\zeta(\eta)d\eta}\hat{T}e^{-i\int_\ell^{\zeta}H^{\text{D}}_{\text{I}}(\eta)e^0_{\zeta}(\eta)d\eta}  |\Psi(\ell)\rj^{\text{S}}
\end{align}
Hence we have expression of Schr\"{o}dinger picture evolution operator using only Dirac picture operators
\begin{align}
U^{\text{S}}(\zeta,\ell)
=&\hat{T}^{-1}e^{-i\int_\ell^{\zeta}H_0^{\text{D}}(\eta)e^0_\zeta(\eta)d\eta}\hat{T}e^{-i\int_\ell^{\zeta}H^{\text{D}}_{\text{I}}(\eta)e^0_{\zeta}(\eta)d\eta}  
\end{align}
and
$
| \Psi(\zeta)\rj^{\text{S}}
=U^{\text{S}}(\zeta,\ell)|\Psi(\ell)\rj^{\text{S}}.
$
We have the transition amplitude  \cite[p.484]{Merzbacher1998}  
\begin{align}
&^{\text{S}}_0\lj B;0| U^{\text{S}}(0,\ell)|A;\ell\rj_0^{\text{S}}\nonumber\\
=&\,^{\text{S}}_0\lj B;0|\hat{T}e^{-i\int_\ell^{0}H_0^{\text{S}}(\eta)e^0_\zeta(\eta)d\eta}\hat{T}e^{-i\int_\ell^{0}H^{\text{D}}_{\text{I}}(\eta)e^0_{0}(\eta)d\eta} |A;\ell\rj_0^{\text{S}}
\end{align}
This is the basis for perturbational calculations since the second factor can be expanded in terms of powers of coupling constant $\lambda$.  In this relation, dependence of fields in  $H^{\text{D}}_{\text{I}}$ on time is the same as in the Heisenberg fields in the non-interacting  case while $H^{\text{S}}_0$ is the same as in Schr\"{o}dinger fields in the non interacting case. $H^{\text{D}}_{\text{I}}$ is supposed to be expressed in terms of $b^\dag_{\bk}(\zeta),  b_{\bk}(\zeta)$ while $H^{\text{S}}$  in terms of $b^{\text{S}\dag}_{\bk}(\zeta),  b^{\text{S}}_{\bk}(\zeta)$.  These two set of quasi-particles operators are related to $a_{\bk},a^\dag_{\bk}$. 
\section{Discussions}
General relativity and QFT are two pillars of modern theoretical physics.  As a preamble of a complete unified quantum theory of  gravity and matter system, quantum field theories in classical curved spacetimes have long been called for. 
In this paper, we proposed a generally covariant framework for quantizing real Klein-Gordon field in de Sitter spacetime. The framework is formulated in conformal coordinate which is specifically chosen. It can be transformed into other coordinate systems $x'$. The fundamental solutions will still be labelled by quantum numbers $\underline{k}$ but the functions will take a more complex appearance depending on the coordinates $x'$. The surfaces $\Sigma$ will be defined by functions $\zeta=\zeta(x')=\text{Const.}$  In the new coordinate system $x'$, the {\it time-dependence} becomes actually $\Sigma$-{\it dependence}.

It is found this framework provides many quantum concepts in parallel with the standard quantum field theories in Minkowski spacetime. The key ingredient for the sake of general covariance is the introduction of vierbein, which furnishes the shift from local coordinate system to tangent space.  It is well-known that concepts such as particle generation and annihilation, particle states, vacuum states, transition amplitude are very important for quantum theory to explain experiments. The vitality of each physical theory lies in its explanatory power as well as predictive power.  The framework proposed in this work is no exception.   Primarily, we obtained a reasonable expression of  measurable energy and momentum. There are many other topics within this framework to be discussed, topics such particle generation and perturbative corrections. 

 Our framework also enjoys the three traditional pictures: Heisenberg, Schr\"{o}dinger and Dirac in an extended fashion. The Hamiltonians in Heisenberg and Schr\"{o}dinger pictures are not identical anymore and  so are not the non-interacting Hamiltonians of Dirac and Schr\"{o}dinger picture equal.  Yet, we can nevertheless devise a way to calculate perturbatively the impact of interaction provided the coupling is weak.
 
Though de Sitter spacetime is of de Sitter symmetry, it is yet to be investigated whether the symmetry can be realized by quantized fields per se. 
The generators of de Sitter algebra $\xi_{AB}=z_A\p_{z^B}-z_B\p_{z^A}$ satisfy
\beq
[\xi_{AB}, \xi_{MN}]=\eta_{BM}\xi_{AN}+\eta_{AN}\xi_{BM}-\eta_{AM}\xi_{BN}-\eta_{BN}\xi_{AM} \label{dSalgebra}
\eeq
 Since at any instant of time $\zeta$, the space is the same as that of Minkowski spacetime and one can directly find operators
\begin{align}
   \hat{\xi}_{ij}(\zeta):&=\sum_{\bp} b^\dag_{\bp}(\zeta)(p_i\p_j-p_j\p_i)b_{\bp}(\zeta)
\end{align}
realize the spatial part of de Sitter algebra. We do not know at this moment whether the full de Sitter algebra can be realized by field operators since there might be quantum symmetry breaking.

Lastly, we expect this framework can be applied to quantization of spinor fields , vector fields in de Sitter space as well. Yet  for other spacetimes such as Robert-Walker spacetime, application of this framework might entail some additional steps.  The difficulty in the Robert-Walker case is that the known basis solutions 
$\Pi^{(\pm)}_{kJ}(\chi)Y^M_J(\theta,\varphi), \underline{k}=(k,J,M)$,
where $\theta,\varphi$ are angular coordinates\cite{Birrell}, are not "plane-waves" anymore, as in de Sitter spacetime. Hence, extra effort is needed to associate the basis solutions to "free particles".   In general, we have to find an appropriate coordinate system in which we can find a complete  set of  basis solutions, which can be interpreted as (or associated to)  "free particles".  Once this is achieved, the framework can be transformed to any other coordinate system and thus made generally covariant.

\end{document}